\pdfoutput=1
\documentclass[%
onecolumn,
preprint,
superscriptaddress,
nofootinbib,
 amsmath,amssymb,
 aps,
floatfix,
]{revtex4-2}
\usepackage{microtype}
\usepackage{graphicx}
\usepackage{tikz}
\usepackage{caption}
\usepackage{subcaption}
\usepackage{color}
\usepackage{microtype}

\usepackage{amsmath}
\usepackage{amssymb}
\usepackage{amstext}
\usepackage{gensymb}
\usepackage{hyperref}
\usepackage{comment}
\usepackage{epsfig}
\usepackage{array}
\usepackage{multirow}
\usepackage{booktabs}

\usetikzlibrary{fit,positioning}
\usetikzlibrary{bayesnet}
\usetikzlibrary{arrows}
\renewcommand{\paragraph}[1]{\textbf{#1}}

\usepackage[version=4]{mhchem}
\usepackage[super]{nth}

\newcommand{\um}{$\mu$m }

\newcommand{\IA}{\AA$^{-1}$}

\renewcommand{\vec}[1]{\mathbf{#1}}
\let\oldhat\hat
\renewcommand{\hat}[1]{\oldhat{\mathbf{#1}}}

\begin{document}

% EXAFS for CCAs is hard, man: Challenges and paths forward for measuring short range order in multicomponent alloys
% EXAFS in CCAs: A five-act tragedy of multicomponent solid solutions, secondary phases, and interstitial oxygen
% Enhancing XAFS Analytical Fidelity for Short range order in CCAs
% The challenge of extracting knowledge from EXAFS measurement on complex concentrated alloys
% High entropy EXAFS analysis: paths towards robust short range order measurement in multicomponent alloys
%\title[EXAFS for CCAs]{Challenges and paths forward for measuring short range order in multicomponent alloys with X-ray absorption spectroscopy}
\title{Why is EXAFS analysis for multicomponent metals so hard?  Challenges and opportunities for measuring ordering in complex concentrated alloys using x-ray absorption spectroscopy.}
\date{\today}

% % use \address with amsart and \affiliation with revtex

\author{Howie Joress}
\email{howie.joress@nist.gov}
\thanks{orcid.org/0000-0002-6552-2972}
\affiliation{Material Measurement Laboratory, National Institute of Standards and Technology, Gaithersburg, MD, USA}

\author{Bruce Ravel}
\email{bravel@bnl.gov}
\thanks{orcid.org/0000-0002-4126-872X}
\affiliation{Material Measurement Laboratory, National Institute of Standards and Technology, Gaithersburg, MD, USA}

\author{Elaf Anber}
\email{eanber1@jhu.edu}
\thanks{orcid.org/0000-0001-7808-7036}
\affiliation{Materials Science and Engineering, Johns Hopkins University, Baltimore, MD, USA}

\author{Jonathan Hollenbach}
\email{jhollen3@jhu.edu}
\thanks{orcid.org/0000-0002-1877-6021}
\affiliation{Materials Science and Engineering, Johns Hopkins University, Baltimore, MD, USA}

\author{Debashish Sur}
\email{ds8vw@virginia.edu}
\thanks{orcid.org/0000-0002-7954-9949}
\affiliation{Department of Materials Science and Engineering, University of Virginia, Charlottesville, VA, USA}

\author{Jason Hattrick-Simpers}
\affiliation{Department of Materials Science and Engineering, University of Toronto, Toronto, ON, Canada}
\email{jason.hattrick.simpers@utoronto.ca}
\thanks{orcid.org/0000-0003-2937-3188}

\author{Mitra L. Taheri}
\email{mtaheri4@jhu.edu}
\thanks{orcid.org/0000-0001-5349-1411}
\affiliation{Materials Science and Engineering, Johns Hopkins University, Baltimore, MD, USA}

\author{Brian DeCost}
\email{brian.decost@nist.gov}
\thanks{orcid.org/0000-0002-3459-5888}
\affiliation{Material Measurement Laboratory, National Institute of Standards and Technology, Gaithersburg, MD, USA}

\begin{abstract}
Short range order is a critical driver of properties (e.g. corrosion resistance and tensile strength) in multicomponent alloys such as complex concentrated alloys (CCAs). 
Extended x-ray absorption fine structure (EXAFS) is a powerful technique well suited for quantifying this short range order.Here, we described in detail  the characteristics of CCAs that make the already challenging task of analyzing EXAFS data even more difficult.
% Further, we identify and analyze sources of model degeneracy that lead to physically distinct but equally good EXAFS fits to CCA data
We then illustrate novel paths towards robust and scalable quantitative SRO analysis which will accelerate the scientific understanding and development of CCAs. 
\end{abstract}

\maketitle

\section{Introduction}

Complex concentrated alloys (CCAs), including multi principal element alloys (MPEAs) and high entropy alloys (HEAs), have attracted significant interest in recent years for their large design space and unique properties\cite{gorsse2018high,miracle2017critical}. 
Many of these desirable properties observed in CCAs have been attributed to local variations in atomic order, which is commonly called short-range order (SRO, technically, SRO includes topological disorder, but here we refer more specifically to chemical short range order). 
SRO has been found to be critical for tailoring many functional properties, including oxidation resistance~\cite{liu2018effect}, mechanical properties~\cite{feng2017effects,inoue2021direct,ding2018tunable}, aqueous corrosion resistance ~\cite{xie2021percolation,tailleart2012effect}, catalytic efficiency~\cite{xiong2019revealing},  magnetic properties~\cite{feng2017effects}, and thermodynamic properties~\cite{chen2022short}. 
For example, local ordering of the first coordination shell can delay the passivation process during corrosion by
reducing the nominal concentration of passivating species required to reach the percolation threshold~\cite{liu2018effect}.
Magnetic properties are similarly affected by SRO:
\textcite{feng2017effects} report a reduction in the atomic magnetic moments of magnetic elements in \ce{FeCoNi(AlSi)_x} when Ni-Al, Co-Si, Fe-Co, Ni-Si, and Fe-Si bonding is increased and Al-Al, Al-Si, and Si-Si pairs are decreased.

%\textbf{Some related literature on SRO: ~\cite{Rao2022}, ~\cite{Zhang2022}, ~\cite{CutsailIII2022}}

SRO is generally characterized using a set of Warren-Cowley (WC) parameters, $\alpha^r_{ij}$, that normalize the number of neighboring atom pairs to the bulk composition in a pairwise fraction\cite{cowley1965short}.
Specifically 
\begin{equation}
\alpha^r_{ij}=1-\frac{P^r(i|j)}{c_i}
\label{eq:WCalpha}
\end{equation}
where the conditional probability $P^r(i|j)$ is the average fractional occupancy of $i$ atoms in the \textit{r\textsuperscript{th}} nearest neighbor (NN) shell around $j \neq i$ atoms and $c_i$ is the fractional concentration of $i$ in the alloy. 
Negative WC values indicate a greater number of $i,j$ pairs than expected in a random solid solution, while positive values indicate fewer $i,j$ pairs than random.

There are multiple methods for experimentally measuring SRO, reviewed in Ref.~\onlinecite{zhang2022characterization}.  
Diffraction based approaches to measuring WC parameters -- by single crystal x-ray, neutron, or TEM diffraction or by x-ray or neutron powder diffraction (including total scattering methods such as pair distribution function (PDF) -- are possible.  
For complex systems, analysis and extraction of SRO parameters is generally accomplished through reverse Monte Carlo (RMC) modeling~\cite{McGreevy2001,Krayzman2009,DiCicco2022} followed by analysis of the simulated volume.
More direct analysis is possible for binary solutions, but this becomes intractable for higher order solutions~\cite{warren_2014}.  
Atom probe tomography can also directly generate the 3D spatial distribution of atoms needed for SRO analysis~\cite{inoue2021direct}. 
One of the strengths of RMC is the joint modeling of EXAFS and PDF observation~\cite{Krayzman2009,DiCicco2022}, which can be particularly useful for constraining the model based on long range information not accessible to EXAFS.
In practice, RMC analysis for EXAFS may be sensitive to initial estimates of non-structural parameters, and systems with complex multiphase structure may require non-trivial post-analysis to summarize the SRO.

Extended x-ray absorption fine structure (EXAFS) and its TEM counterpart extended energy loss fine structure (EXELFS, described in \S~\ref{subsec:exelfs})\cite{hart2023revealing} are spectroscopic techniques that use the scattering of a photoelectron to infer local chemical information in the $\approx 5 $ \AA{} about an absorbing atom.   
EXAFS analysis, as described in \S\ref{subsec:complexity}, extracts the pairs of NN occupancies that, along with sample composition, can be used to directly calculate WC parameters.  
EXAFS has several advantages over other techniques. 
First, the probed volume is very large ($\approx$mm\textsuperscript{2} area with \um scale depth sensitivity) compared to TEM and atom probe, providing ensemble averaged values.  
Second, the way EXAFS is measured inherently speciates the absorbing atom, producing signals that are sensitive to a set partial pair-distribution functions rather than a single PDF generated by x-ray total scattering.   
This is particularly advantageous for minority species which are common and play key roles in CCAs.  
Third, EXAFS analysis directly parameterizes the values needed for calculating the WC parameters, particularly in cases where multiple phases or partial long range order need to be explicitly modeled. 
Finally, the sample preparation requirements for EXAFS are relatively modest and non-invasive, requiring thinning to microns instead of nanometers, enabling this technique to be readily scalable for a large number of samples or conditions, with minimal risk for artifacts induced by sample preparation.
These minimal requirements also make EXAFS ideal for \textit{in situ} experiments, such as heating.

While EXAFS analysis is a powerful tool for characterization of SRO, quantitative analysis is notoriously difficult~\cite{Calvin2013,kido2020problems,ravel2007}.
%Reliably measuring coordination numbers is even more so~\cite{ravel2007}, even with well specified physical models.
%The two principal reasons for the difficulty of EXAFS analysis are the limited information content in the spectra~\cite{Stern1993} and the high degree of degeneracy between structural parameters.
%Typical EXAFS models are highly overparameterized, and in our experience analyzing CCAs, it is possible to find multiple defensible solutions, even if all parameters optimize to physically defensible values.  
%There are multiple methods to measure SRO experimentally, such as TEM-based diffraction techniques and PDF [?], but Extended X-ray Absorption Fine Structure (EXAFS) is a powerful method for determining chemical short-range order and local structures in materials. 
% Beyond making measurements and Warren-Cowley parameterization of single samples possible, we are interested in enabling high-throughput EXAFS analysis of alloys of this class possible, including semiautomated quantification.
%\paragraph{There are multiple methods for quantifying SRO, but EXAFS is best/powerful}
%\paragraph{specific challenges for CCAs}

\begin{figure}[h!]
    \centering
    \includegraphics[width=165mm]{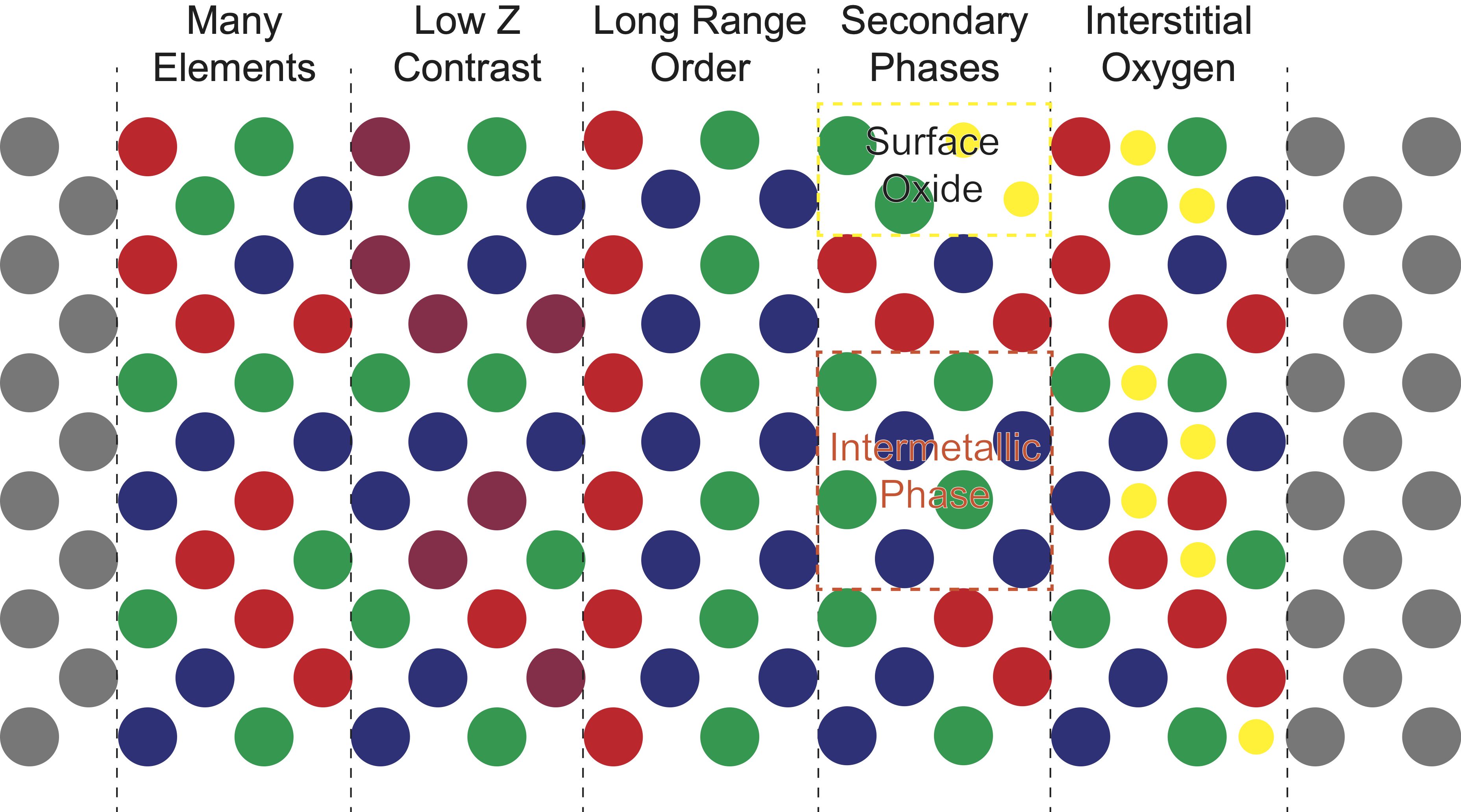}
    \caption{A schematic demonstrating five aspects of CCAs that complicate EXAFS analysis.  The red, blue, green, purple, and grey circles represent metal atoms.  The yellow circles represent oxygen atoms.}
    \label{fig:tldr}
\end{figure}
For CCAs there are many compounding effects that make this particularly challenging. In this perspective we outline five aspects of CCA systems, illustrated in Fig.~\ref{fig:tldr}, that make them uniquely challenging for quantitative SRO analysis via EXAFS:
\begin{enumerate}
\item Multicomponent solid solutions introduce a large number of parameters
%\item interference at higher shells reduces spectral weight, which limits ability to share information across the fuller spectra
\item CCAs often have many elements with similar atomic number (Z), leading to poor elemental contrast (\S\ref{subsec:contrast})%, limiting questions accessible by analysis
\item Potential long range order (\S\ref{sec:LRO}) dramatically increases the complexity of the structural models
\item Secondary phases, both metallic and oxide, contribute the to EXAFS signal and must be included in the model (\S\ref{sec:secondphase}).
\item Interstitial oxygen is present at significant concentrations in many alloys, particularly refractory CCAs, contributing to the scattering signal below the first metal-metal NN distance (\S\ref{sec:secondphase}).
\end{enumerate}

We discuss these challenges in detail and identify paths towards robust and scalable quantitative SRO analysis to accelerate scientific understanding and development of complex, technologically relevant CCA systems.

A detailed overview of EXAFS is beyond the scope of this paper. 
There are many excellent resources for this including Refs. \onlinecite{Calvin2013,newville2014fundamentals}

% what are the limitations that present roadblocks for multi-component EXAFS fitting"
\section{Current limitations of EXAFS analysis for multicomponent alloys}\label{sec:challenges}
\subsection{The EXAFS equation and its complexity}\label{subsec:complexity}
\begin{figure}[h!p]
     \centering
     \begin{subfigure}[b]{48mm}
         \centering
          \includegraphics[width=.9\textwidth]{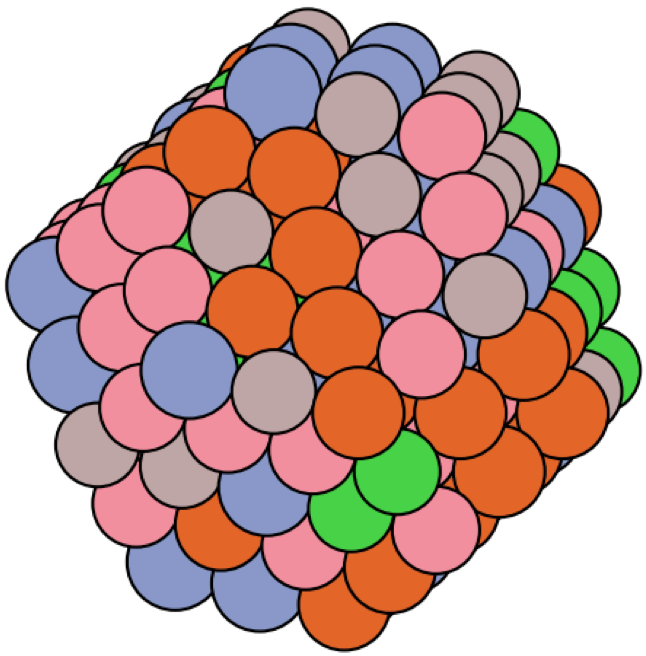}
         \caption{}
         %\label{fig:cluster}
     \end{subfigure}
     %\hfill
     \begin{subfigure}[b]{48mm}
         \centering
         \includegraphics[width=.9\textwidth]{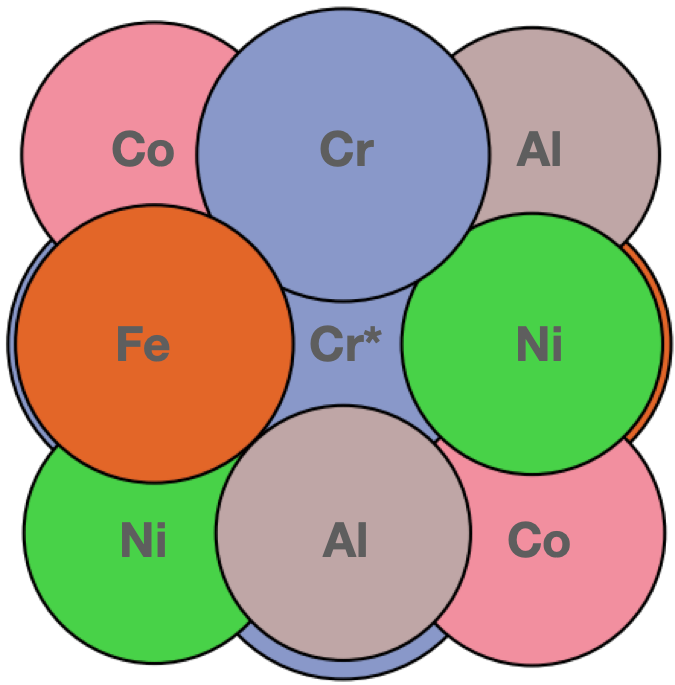}
         \caption{}
         \label{fig:firstshell}
     \end{subfigure}
     %\hfill
     \begin{subfigure}[b]{65mm}
         \centering
         \includegraphics[width=\textwidth]{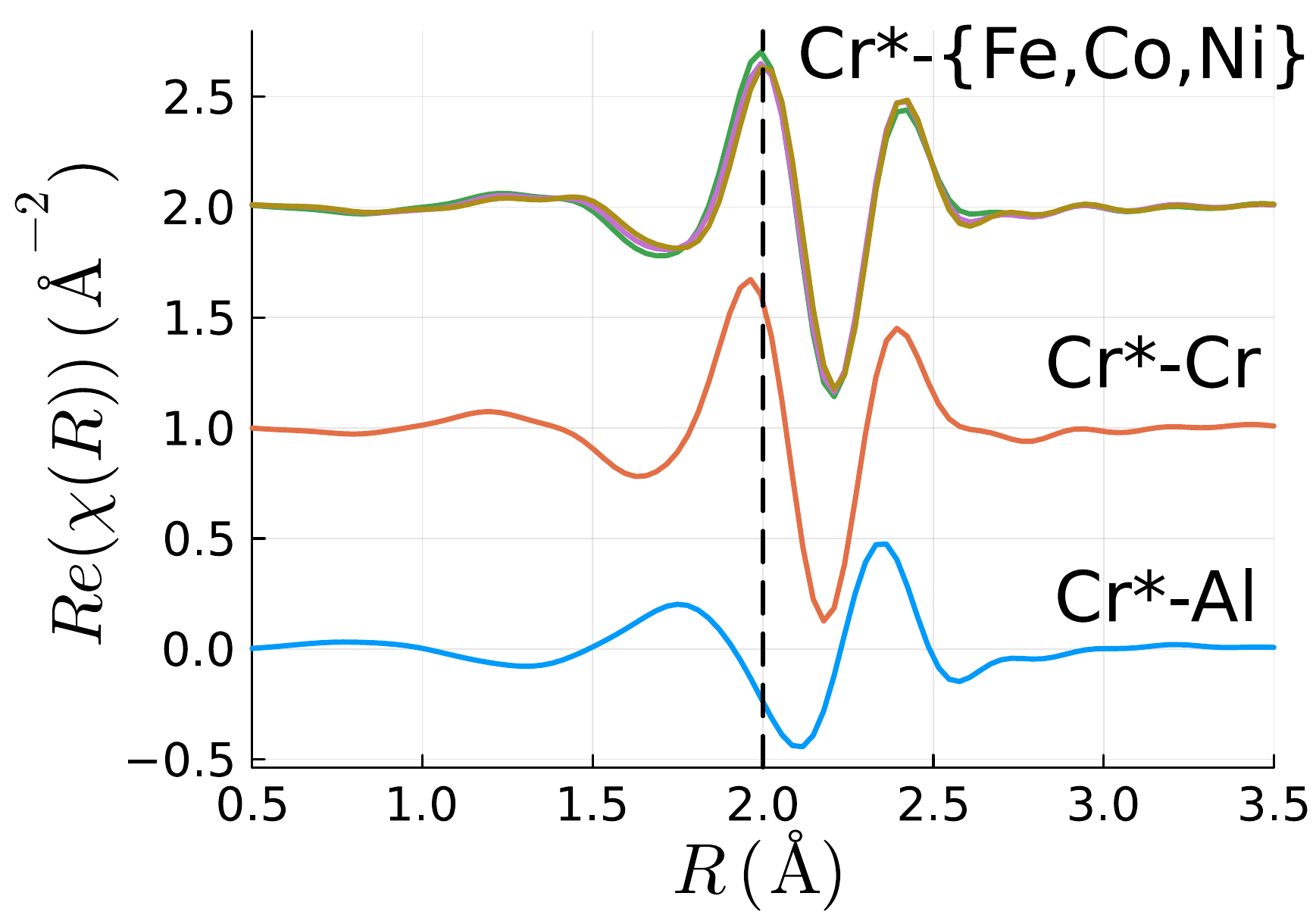}
         \caption{}
         \label{fig:pathexpans}
     \end{subfigure}
        \caption{(a) Quinary disordered FCC solid solution atomic cluster for FeCoNiCrAl (b) Cr first coordination shell used to parameterize EXAFS model (c) The real part of the EXAFS contribution from each scattering path in (b) (vertical offset for clarity). This figure illustrates the low degree of contrast between neighboring atoms, with Fe, Co, and Ni scatters grouped together.  The vertical dashed line acts as a guide to the eye.  The asterisk (*) represents the absorbing species.}
        \label{fig:fig1}
\end{figure}

% While there are opportunities for physically-motivated parameter constraints, in practice they can be difficult to implement.
The EXAFS path expansion equation models the EXAFS signal, $\chi(k)$, of the average absorbing atom as the sum of the contribution from a set of photoelectron scattering paths, corresponding to each neighboring species or sets of species.
The structural model for even simple CCAs without secondary phases or long range order contains a large diversity of scattering paths, resulting in unfavorable model size scaling with the number of species.
Figure~\ref{fig:fig1} shows a quinary FeCoNiCrAl FCC (Face Centered Cubic) nanocluster (a), along with (b) the first coordination shell surrounding a single symmetrically distinct absorbing Cr atom in the FCC solid solution.
This highlights two intrinsic difficulties of EXAFS analysis for CCAs.
First, even a simple first neighbor shell model has high complexity from the many distinct scattering paths.
Consideration of higher order NN shells requires explicit modeling of multiple scattering paths, which scale combinatorially with the number of scattering elements.
This limits almost all practical analysis of SRO CCAs to the first or first and second shell (with exceptions, \textit{e.g.,} Ref. \cite{ravel2002exafs}).
We discuss the trade-offs of this approach in \S\ref{subsec:highR}, but otherwise limit our discussion to first shell analysis.
The second issue is the low contrast between the scattering factors of neighboring elements as shown in ~\ref{fig:fig1}c and discussed in \S\ref{subsec:contrast}.
%More concerning, for alloy systems with many species of similar atomic number the scattering signals are remarkably similar (\textit{e.g.,} the $Cr^*-\{Fe,Co,Ni\}$ paths shown in Fig~\ref{fig:SRO-model-cartoon}c).
%We discuss in detail the problem of scattering signal degeneracy in Section~\ref{sec:degeneracy} and \ref{sec:Zcontrast}.

Consider the functional form of the EXAFS path expansion equation,

\begin{equation}
    \label{fig:exafseq}
    \chi(k) = S_0^2 \sum_{\gamma=1}^{\Gamma} N_\gamma \frac{f_\gamma(k)}{kd_\gamma^2} e^{\frac{-2d_\gamma}{\lambda_\gamma(k)}} e^{-2k^2\sigma_\gamma^2} \sin \left(2kd_\gamma + \delta_\gamma(k) \right),
\end{equation}
where $\chi$ is a function of wavenumber $k$ of the generated photoelectron
\footnote{$k=\sqrt{2m_eE/\hbar^2}$ where $E$ is the energy of the photoelectron ($E-E_{0})$}
, parameterized by --- for each scattering path $\gamma$ in the model --- at least five empirical parameters typically estimated by nonlinear least squares and three functions of $k$ obtained through first principles multiple scattering calculations, \textit{e.g.,} FEFF\cite{Zabinsky1995}.
%When considering only single atom scattering paths, a unique path exists for each element in each NN shell. 
The three theory-derived values are $f_\gamma(k)$, $\lambda_\gamma(k)$ and $\delta_\gamma(k)$, the theoretical effective scattering amplitude, photoelectron mean free path (in FEFF this is interpolated from tabulated values), and the phase shift for each scattering path respectively.

% Page 3: "the amplitude reduction factor to normalize the pattern 
% intensity" -- that's not quite right.  S02 is a constant term that 
% approximates the effect of doing the one-electron calculation in the 
% sudden approximation. See   https://doi.org/10.1103/RevModPhys.72.621 
% and https://doi.org/10.1103/PhysRevB.52.6332

% In Artemis, the S02 term is used to accommodate the physical S02 as well 
% as any other things that might effect the amplitude of the signal 
% compared to the calculation.  As such, it is something like a 
% normalization factor.  But to call it that is likely to get flagged by 
% whomever reviews the paper.

The five empirical parameters per scattering path are
\begin{enumerate}
    \item $S_0^2$: the amplitude reduction factor that compensates for the sudden, single-electron assumption made by FEFF;
    \item $\Delta E_0$: change in the expected edge energy (\textit{i.e.,} the zero-point for $k$);
    \item $N_\gamma$: the multiplicity\footnote{In the EXAFS literature $N_\gamma$ is often called the \textit{degeneracy} of the scattering path; in this manuscript we use the term \textit{multiplicity} to avoid confusion during discussion of degenerate least squares optimization problems.} of the $\gamma$ scattering path;
    %For single scattering paths this is often interpreted as a coordination number for that species in a given NN shell. A constraint exists such that $\sum_{\Gamma} N_\gamma$ is equal to the total coordination number of, in dense metals, for that shell. 
    \item $d_\gamma$: half the photoelectron path length, interpreted for single scattering paths as an average bond length. The first principles modeling by FEFF requires a starting path length, $d_\gamma^0$ and fitting is performed on $\Delta\ d_\gamma$ such that $d_\gamma=d_\gamma^0+\Delta\ d_\gamma$;
    \item$\sigma^2_\gamma$: the variance of the half-path length, often called the 
    %Debye-Waller factor or 
    mean square relative displacement (MSRD) and sometimes conflated with the Debye-Waller factor, which is a portion thereof;
\end{enumerate}

In practice, $S_0^2$ accommodates the physical $S_0^2$ as well as any other factors that might affect the amplitude of the signal compared to the calculation, for example experimental geometry, sample inhomogeneity, detector non-linearity, and beam harmonic content.

For the case of crystalline solid solutions, we can rewrite Eq.~\ref{fig:exafseq} as a double sum,
explicitly organizing the $\gamma$ paths into $R$ sets of neighbor shells $r$ each with $M$ scattering elements $m$:
\begin{equation}
    \label{fig:exafseq-shell}
    \chi(k) = S_0^2 \sum_{r=1}^{R}\sum_{m=1}^M N_{r,m} \frac{f_{r,m}(k)}{kd_{r,m}^2} e^{\frac{-2d_{r,m}}{\lambda(k)_{r,m}}} e^{-2k^2\sigma_{r,m}^2} \sin \left(2kd_{r,m} + \delta_{r,m}(k) \right).
\end{equation}
Strictly speaking, considering multiple scattering this is a set of geometrically identical paths. 
Crystalline materials without vacancies follow the constraint that $\sum_m^MN_{r,m}$ is equal to the total coordination number of the $r^{th}$ shell.
Division by this coordination number yields the conditional probability for the  $r^{th}$ shell needed to derive the WC parameter: $P^r(i|j)=N_{r,i}/{\sum_m^MN_{r,m}}$ when measured about a $j$ absorber ($j\neq i$).  
As we will be focusing on first shell analysis, notationally we will assume $r=1$ and $\gamma=m$ or drop the indexes entirely.

\begin{figure}[h!btp]
    \centering

    \begin{subfigure}[b]{51mm}
        %%%%%%%%%%%%%%%%%%%%%%%%%%%%%%%%%%%%%%%%%%%%%%%%%%%%%%%%%%%%%%%
%
% Welcome to Overleaf --- just edit your LaTeX on the left,
% and we'll compile it for you on the right. If you open the
% 'Share' menu, you can invite other users to edit at the same
% time. See www.overleaf.com/learn for more info. Enjoy!
%
%%%%%%%%%%%%%%%%%%%%%%%%%%%%%%%%%%%%%%%%%%%%%%%%%%%%%%%%%%%%%%%
% % \documentclass{standalone}
% \usepackage{tikz}
% \usepackage{graphicx}
% \usepackage{caption}
% \usepackage{subcaption}
% \usepackage{color}

% \usetikzlibrary{fit,positioning}
% \usetikzlibrary{bayesnet}
% \usetikzlibrary{arrows}
% \begin{document}
%\begin{adjustbox}
\begin{tikzpicture}

    % observed profile
     \node[obs] (y) {$\chi_i$};%
     \node[det,left=of y,yshift=.5cm] (mu) {$\chi_i'$}; %
     \node[latent,left=of y,yshift=-0.5cm] (var) {$\epsilon_i$}; %

     % path variables
     \node[latent,above=1.4cm of mu,xshift=-0.9cm] (msrd) {$\sigma^2_\gamma$}; %

     \node[det,above=1.3cm of mu,xshift=1cm] (N) {$N_\gamma$}; %

     \node[latent, above=1.4cm of mu](r){$d_\gamma$};
     \node[const,left=1.13cm of mu, yshift=.5cm] (S02) {$S_0^2$}; %
     \node[latent,left= of mu, yshift=-.5cm,xshift=0.1cm] (E0) {$\Delta E_0$}; %

    \node[latent,above=of N,yshift=-.5cm](Nhat) {$\vec{N}$};

    % plate
     \plate[inner sep=0.3cm] {profile} {(y) (mu) (var)} {$I$}; %
     \plate[inner sep=0.3cm] {paths} {(msrd) (r) (N) } {$\Gamma$}; %

    % edges
     \edge {var, mu} {y}
     \edge {msrd, E0, N, S02,r} {mu}
     \edge {Nhat} {N}
    %  \edge {a, fraction, scale} {amp}
    %  \edge {U, V, W, mus, sz} {sigma}
    %  \edge {X, Y, Z, mus, sz} {gamma}
    %  \edge {sz_mean, sz_var} {sz}
    %  \edge {gp_s, gp_ls} {fraction}

\end{tikzpicture}

%\end{adjustbox}
% \end{document}
        \subcaption[]{}
        \label{fig:model-diagrams:a}
    \end{subfigure}
    \hspace{1cm}
    \begin{subfigure}[b]{51mm}
        %%%%%%%%%%%%%%%%%%%%%%%%%%%%%%%%%%%%%%%%%%%%%%%%%%%%%%%%%%%%%%%
%
% Welcome to Overleaf --- just edit your LaTeX on the left,
% and we'll compile it for you on the right. If you open the
% 'Share' menu, you can invite other users to edit at the same
% time. See www.overleaf.com/learn for more info. Enjoy!
%
%%%%%%%%%%%%%%%%%%%%%%%%%%%%%%%%%%%%%%%%%%%%%%%%%%%%%%%%%%%%%%%
% \documentclass{standalone}
% \usepackage{tikz}
% \usepackage{graphicx}
% \usepackage{caption}
% \usepackage{subcaption}
% \usepackage{color}

% \usetikzlibrary{fit,positioning}
% \usetikzlibrary{bayesnet}
% \usetikzlibrary{arrows}
% \begin{document}

\begin{tikzpicture}

    % observed profile
     \node[obs] (y) {$\chi_i$};%
     \node[det,left=of y, yshift=0.5cm] (mu) {$\chi_i'$}; %
     \node[latent,left=of y, yshift=-0.5cm] (var) {$\epsilon_i$}; %

     % path variables
     \node[det,above=1.35cm of mu,xshift=-1.3cm] (msrd) {$\sigma^2_{a,\gamma}$}; %
     
     \node[det,above=1.35cm of mu,xshift=1.3cm] (N) {$N_{a,\gamma}$}; %

     \node[det, above=1.4cm of mu](r){$d_{a,\gamma}$};
     \node[,left=1.13cm of mu, yshift=.5cm] (S02) {$S_0^2$}; %
     \node[latent,left= of mu, yshift=-.5cm] (E0) {$\Delta E_0$}; %

    \node[latent,above=of N, yshift=-.5cm](Nhat) {$\vec{N}$};     
    \node[latent, above=of msrd, yshift=-.4cm](Sigma){$\Sigma^2$};
    \node[latent, above=of r, yshift=-.4cm](R){$D$};

    % plate
     \plate[inner sep=0.3cm] {profile} {(y) (mu) (var)} {$I$}; %
     \plate[inner sep=0.3cm] {paths} {(msrd) (r) (N) } {$\Gamma(path)$}; %
     \plate[inner sep=.1cm] {edge} {(paths) (profile) (E0) (S02)}{$A(absorber)$};

    % edges
     \edge {var, mu} {y}
     \edge {msrd, E0, N, S02,r} {mu}
     \edge {Nhat} {N}
      \edge {Sigma} {msrd}
      \edge {R} {r}
    %  \edge {X, Y, Z, mus, sz} {gamma}
    %  \edge {sz_mean, sz_var} {sz}
    %  \edge {gp_s, gp_ls} {fraction}

\end{tikzpicture}

%\end{document}
        \subcaption[]{}
        \label{fig:model-diagrams:b}
    \end{subfigure}    
    \\ \vspace{0.5cm}
    \begin{subfigure}[b]{112mm}
        %%%%%%%%%%%%%%%%%%%%%%%%%%%%%%%%%%%%%%%%%%%%%%%%%%%%%%%%%%%%%%%
%
% Welcome to Overleaf --- just edit your LaTeX on the left,
% and we'll compile it for you on the right. If you open the
% 'Share' menu, you can invite other users to edit at the same
% time. See www.overleaf.com/learn for more info. Enjoy!
%
%%%%%%%%%%%%%%%%%%%%%%%%%%%%%%%%%%%%%%%%%%%%%%%%%%%%%%%%%%%%%%%
% \documentclass{standalone}
% \usepackage{tikz}
% \usepackage{graphicx}
% \usepackage{caption}
% \usepackage{subcaption}
% \usepackage{color}

% \usetikzlibrary{fit,positioning}
% \usetikzlibrary{bayesnet}
% \usetikzlibrary{arrows}
% %\begin{document}

\begin{tikzpicture}

    % observed profile
     \node[obs] (y) {$\chi_i$};%
     \node[det,left=of y, yshift=0.5cm] (mu) {$\chi_i'$}; %
     \node[latent,left=of y, yshift=-0.5cm] (var) {$\epsilon_i$}; %

     % path variables
     \node[det,above=1.3cm of mu,xshift=-1.1cm] (msrd) {$\sigma^2_{\gamma}$}; %
     
     \node[det,above=1.3cm of mu,xshift=1.1cm] (N) {$N_{\gamma}$}; %

     \node[det, above=1.35cm of mu](r){$d_{\gamma}$};
     \node[const,left=2.5cm of mu, yshift=.5cm] (S02) {$S_0^2$}; %
     \node[latent,left= of mu, yshift=-.5cm] (E0) {$\Delta E_0$}; %

    \node[det,right=of N, yshift=0cm](Nhat) {$\vec{N}_s$};     
    \node[latent,above=of Nhat, yshift=-.4cm](Ngp){$\vec{N}_{gp}$};
    \node[latent, above=of msrd, yshift=-.5cm](siggp){$\sigma^2_{gp}$};
    \node[latent, above=of r, yshift=-.4cm](rgp){$d_{gp}$};

    % plate
     \plate[inner sep=0.3cm] {profile} {(y) (mu) (var)} {$I$}; %
     \plate[inner sep=0.3cm] {paths} {(msrd) (r) (N) (siggp)(rgp)} {$\Gamma(path)$}; %
     \plate[inner sep=.3cm] {edge} {(profile) (E0) (msrd)(r)(N)(Nhat)}{$S(sample)$};

    % edges
     \edge {var, mu} {y}
     \edge {msrd, E0, N, S02,r} {mu}
     \edge {Nhat} {N}
     \edge{Ngp}{Nhat}
     \edge {siggp} {msrd}
      \edge {rgp} {r}
    %  \edge {X, Y, Z, mus, sz} {gamma}
    %  \edge {sz_mean, sz_var} {sz}
    %  \edge {gp_s, gp_ls} {fraction}

\end{tikzpicture}

%\end{document}
        \subcaption[]{}
        \label{fig:model-diagrams:c}
    \end{subfigure}
    
    \caption{(a) Plate diagram illustrating a standard, single edge EXAFS fit.  An independent multi-edge fit would add an additional plate around the entirety of the diagram for each absorber. (b) A multi-edge model with self-consistency assumptions. (c) A multi-sample multi-edge model with both self-consistency and smoothness assumptions. Variables are defined in the text.}
    \label{fig:model-diagrams}
\end{figure}
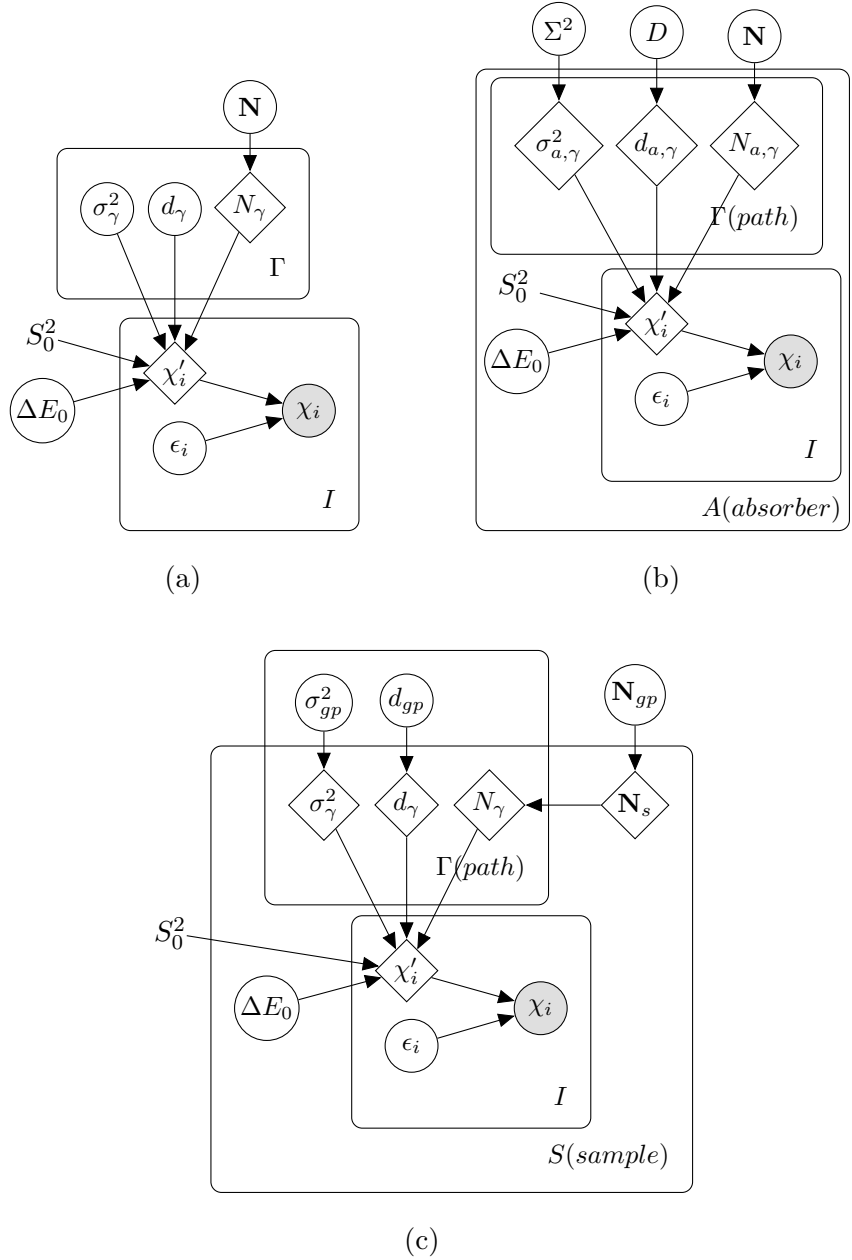

Fig.~\ref{fig:model-diagrams} illustrates the relationship between model parameters under various assumptions.
These plate diagrams are a convenient graphical tool for concretely and concisely representing conditional dependence relationships between variables in complex graphical models~\cite{buntine1994operations}. 

% to make this fit a column, set the subfigure withs to 112mm and separate with \\

Fig.~\ref{fig:model-diagrams:a} illustrates a standard single-edge EXAFS model with no simplifying assumptions. 
The rectangular ``plates'' show collections of variables that are conditionally independent given their ancestors in the graph.
Circular nodes represent independent parameters in the model, directed edges (arrows) represent conditional dependence relationships, and diamond nodes represent variables that are deterministic given the values of their ancestors in the graph.
Filled nodes represent observed variables, for example the measured EXAFS values $\chi_i$ for $I$ values of $k$ (or $R$ in the case of the Fourier transformed signal), which is dependent on the underlying noiseless EXAFS signal, $\chi_i'$, and detector noise, $\epsilon_i$.

The upper plate in the single-edge first shell model of Fig.~\ref{fig:model-diagrams:a} concisely illustrates the EXAFS path expansion model for the $\Gamma$ first shell scattering paths included in the model.
Each scattering path has independent MSRD and bond length parameters, $\sigma^2_\gamma$ and $d_\gamma$.
The coordination variables $N_\gamma$ are shown as deterministic variables because they are coupled by sampling $M-1$ independent parameters that form a simplex variable $\mathbf{N}$.
The absorption edge $\Delta E_0$ is drawn outside the path expansion plate indicating the common assumption that $E_0$ is constant for all scattering paths.
Similarly, the amplitude reduction factor $S_0^2$ is drawn without a node marker, indicating the assumption of a fixed value transferred from analysis of an elemental standard.

In addition to the parameters described above, in some cases, additional path parameters can be introduced, for example  the higher order cumulant expansion method for asymmetric path length distributions~\cite{fornasini2001cumulant}.
Structures with extreme structural disorder may be require modeling the path length distribution with a histogram expansion method~\cite{Price2012}.
With such a large array of structural parameters, most structural models are heavily overparameterized with respect to the limited amount of information contained in EXAFS data~\cite{Stern1993}.

The information theoretic Nyquist criterion~\cite{Stern1993} gives a heuristic bound on the number of independent parameters supported by the data: $N_{ind} \approx \frac{2\Delta k \Delta R}{\pi}$.
$\Delta k$ is the range of k space that is Fourier transformed, generally varying from 6 \IA\ for a quick scan and up to 12 \IA\ for the highest quality data.
$\Delta R$ is the fitting range in real space, typically spanning 1 \AA{} to 2 \AA{} for the first shell in CCA solid solutions.
This leads to a range of $N_{ind}$ from 4 to 15 per spectrum.  

We compare the amount of information based on this criterion with the number of model parameters as a function of alloy components $M$ in the first three columns of Table \ref{tab:overparameterization} (In this table we use $N_{ind}=8$ per edge, as this is a reasonably conservative value and is consistent with the data we generally collect).    
$N_{ind}$ is roughly multiplicative with the number of edges collected and analyzed.
In the independent FCC model described above, there are $3M^2$ independent parameters (assuming a fixed $S_0^2$) if we independently model all $M$ edges; alloys with more than two species are overparameterized.  
We also show in the \nth{4} column parameter counts for body centered cubic (BCC) materials,
where it is not feasible to isolate the first shell with Fourier filtering because of the close overlap of the first and second NN  bond lengths.
Consequently, the number of independent parameters is substantially higher ($6M^2-M$).
%N_ind = 2 * (k = 11 - 2) * (R = 3 - 1.5) / π
The overparameterization of these multicomponent solid solution models leads to particularly challenging EXAFS analysis problems if we must accurately measure chemical SRO.
A minimally constrained first-shell model for an FCC quinary alloy has 15 parameters for a single edge, nearly double what a typical EXAFS spectrum supports.

\begin{table}[]
    \centering
    \begingroup
    \renewcommand{\arraystretch}{1.1}
    \begin{tabular}{l>{\bf}r>{\hspace{2pc}}rr>{\hspace{1pc}}rr>{\hspace{1pc}}rr>{\hspace{1pc}}cc}
    \toprule
    M & $N_{ind}$ & \multicolumn{2}{c}{Independent} & \multicolumn{2}{c}{Symmetric}&\multicolumn{2}{c}{SRO}& Symmetric \& metal radii & SRO \& metal radii \\ 
    \cmidrule(lr){3-4}\cmidrule(lr){5-6}\cmidrule(lr){7-8}\cmidrule(lr){9-9}\cmidrule(lr){10-10}
           &    &    FCC  & BCC &     FCC & BCC&     FCC & BCC & FCC & FCC \\
    \midrule
        1 &  8 & \bf{3}  & \bf{5} &         &     &         &         &     & \\
        2 & 16 & \bf{12} & 22     & \bf{10} & 18  & \bf{9}  & \bf{16} &   \textbf{9} & \textbf{8} \\
        3 & 24 &     27  & 51     & \bf{21} & 39  & \bf{17} & 31      &  \textbf{18} & \textbf{14} \\
        4 & 32 &     48  & 92     &     36  & 68  & \bf{27} & 50      &  \textbf{30} & \textbf{21} \\
        5 & 40 &     75  & 145    &     55  & 105 & \bf{39} & 73      &  45          & \textbf{29} \\
        6 & 48 &     108 & 210    &     78  & 150 &    53   & 100     &  63          & \textbf{38} \\
    \bottomrule
    \end{tabular}
    \endgroup
    \caption{Parameter counts for first shell solid solution models, assuming we can measure and fit all $M$ edges. Bolded entries show systems that are in principle tractable under these assumption sets.  \S \ref{subsec:complexity} describes the independent and symmetric models,   and \S\ref{subsec:multiedge} describes the SRO model.}
    \label{tab:overparameterization}
\end{table}

Whenever EXAFS for multiple absorption edges is available, the precision and accuracy of the analysis can be improved through joint multi-edge analysis by coupling the structural models for each absorbing species with physical self-consistency constraints~\cite{Ravel1999} as shown schematically in Fig. \ref{fig:model-diagrams:b}.
Calvin \textit{et al.}~\citep{Calvin2002} demonstrate an impressive suite of such assumptions for analysis of manganese zinc ferrite nanoparticles.
If we apply this approach in a self-consistent solid solution model, the bond length $d$ and MSRD parameters $\sigma^2$ can be modeled as $M \times M$ symmetric matrices, $D$ and $\Sigma^2$ respectively.
Because the bond lengths and MSRDs between pairs of species must match, only the upper triangular part of this matrix represents free parameters.
For instance, if atoms of scattering species A have some average distance from NN atoms of absorbing species B,  by construction absorbing species A atoms must have the same average distance from NN atoms of species B. 
This symmetry assumption reduces the $M^2$  bond lengths in an independent model to $M+(M^2 - M) / 2$ pairwise bond lengths.
The number of MSRD parameters scales similarly.
The affect of this assumption is illustrated in the ``symmetric'' column of Table \ref{tab:overparameterization}, where the  FCC model has $2M^2 + M$ independent parameters.

Unfortunately, for higher order alloys this assumption is insufficient to eliminate the overparameterization.
This is often address through simplifying assumptions, for example path length and MSRD equality constraints within neighbor shells.
For disordered solid solutions, which commonly exhibit high degrees of static lattice distortion, these assumptions may introduce unacceptable levels of unphysical bias that makes accurate SRO quantification impossible.
We discuss these parameterization issues and potential mitigation strategies in detail in Section~\ref{sec:approach}.

\subsection{Parameter degeneracy}\label{sec:degeneracy}
It is well known in EXAFS analysis that there is a high degree of parameter degeneracy. 
That is, often multiple sets of fitting parameters provide equally optimal goodness of fit\cite{Calvin2013}. 
In particular $\Delta E_{0}$ and $d_\gamma$ are known to have a high degree of correlations because both parameters affect the phase term of the complex scattering factor in the EXAFS equation ~\cite{Kelly2007,michalowicz1998multiple}.  
While element-specific bond lengths are important parameters in these alloys, even more important is the occupancy of elements $N$ in the shell.
Unfortunately $N$ is also highly correlated with other values, all of which affect the amplitude of $\chi$.  
For a single shell, single edge fit, $S_0^2$ is completely degenerate with $N$.  
The MSRD parameter is also highly correlated, but is more separable based on its $k$ dependence.  

\begin{figure*}[h!tbp]
    \centering
    \includegraphics[width=\textwidth]{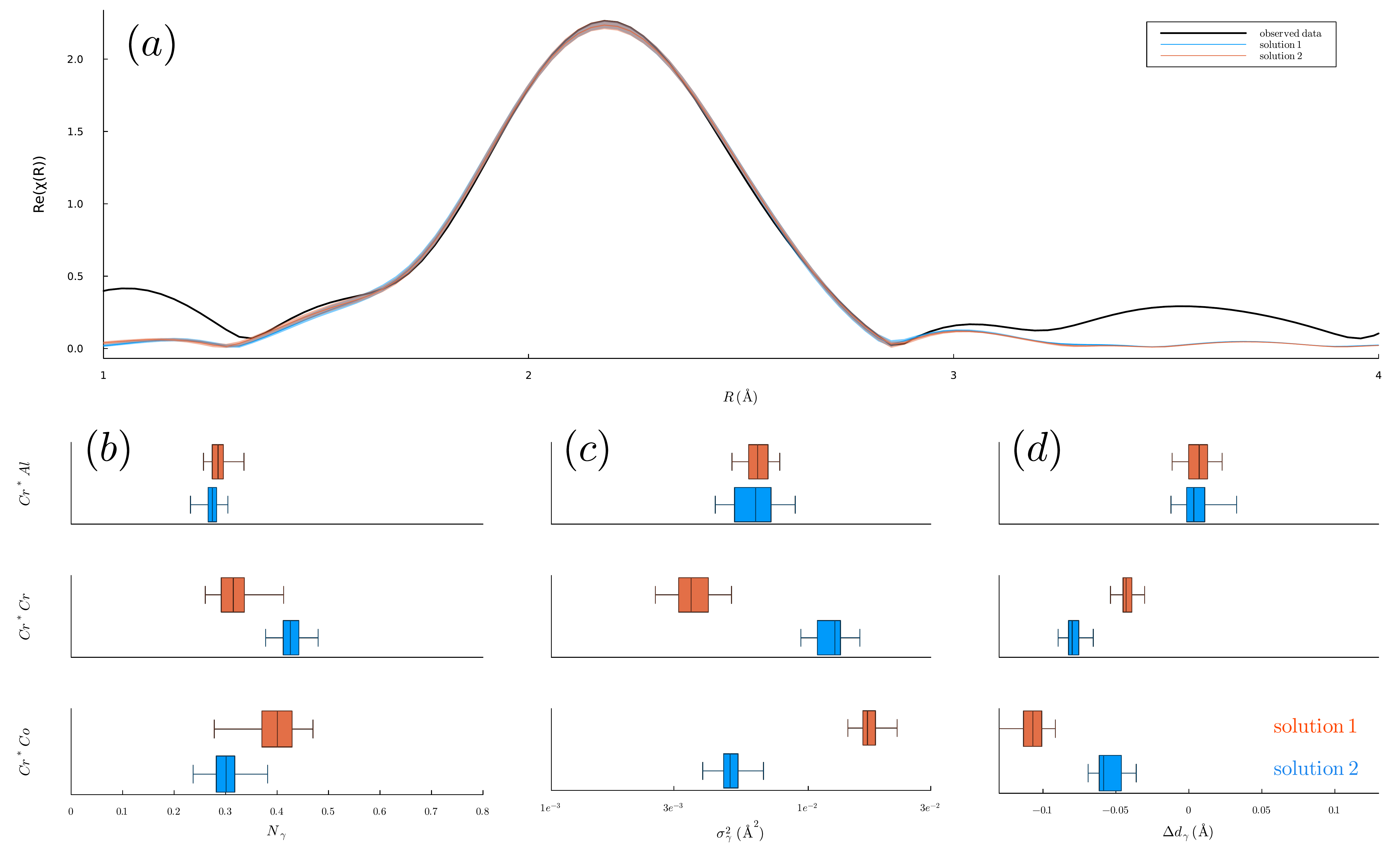}
    \caption{Two degenerate solutions (illustrated in red and blue) using the same physical model to fit BCC CoCrAl EXAFS data taken on the Cr edge.  (a) shows the fits to the real component of the Fourier transformed data and (b-d) shows the parameters for the two models with their associated uncertainties for the NN shell (though the first two shells were modeled). In (b), $N_\gamma$ is the fractional occupancy of the 8 sites in that shell.  Parameter degeneracy in even simplified EXAFS models can be high enough that even physical heuristics for parameter values are not sufficient to constrain analysis to unique solutions in all cases.
    }
    \label{fig:nonidentifiable}
\end{figure*}

Fig. \ref{fig:nonidentifiable} shows some BCC CoCrAl Cr-edge EXAFS data that illustrate this issue.
There are two distinct solutions (illustrated in red and blue) that fit the data equally well despite having substantially different coordination and bonding parameters.
The R factors, $R = \frac{\sum_{i=1}^N (\chi_i - \hat{\chi}_i)^2}{\sum_{i=1}^N \chi_i^2}$, for the two solutions are 0.00362 and 0.00238, indicating excellent consistency with the data.
The edge energy shifts (not shown) are similar at $-7.28 \pm 1.03 $ eV and  $-6.01 \pm 1.35 $ eV, ruling out unphysical compensation of phase shift induced by changing bond lengths and coordination variables~\cite{Kelly2007}.  
 % this is from CoCrAl-analysis-opt
 % maybe need to find the simplified FCC model that was hard to fit...
 % since I think this model still has more parameters than Nyquist would allow...

\begin{figure}
    \centering
    \includegraphics[width=0.9\textwidth]{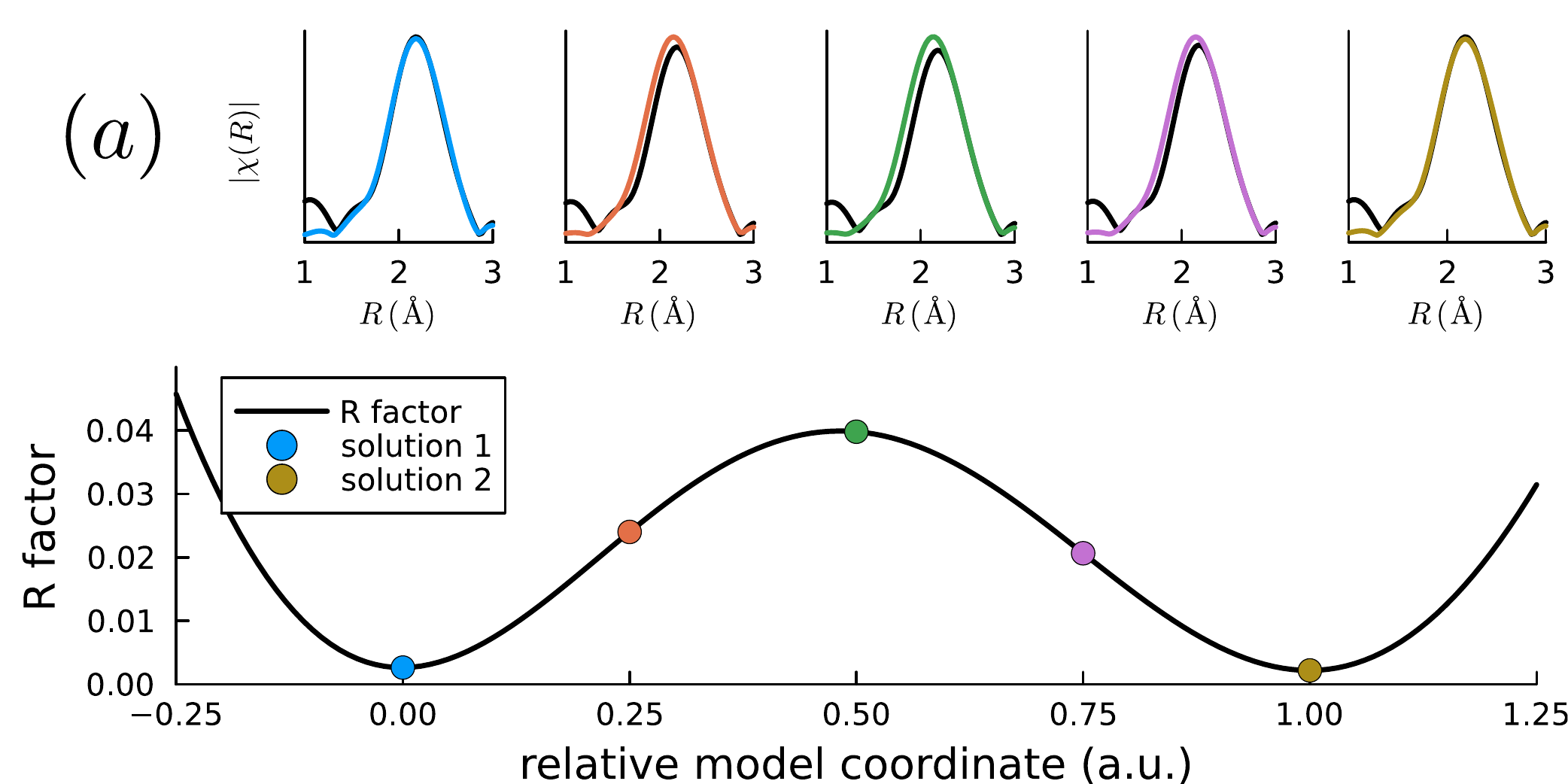} \\
    \includegraphics[width=0.9\textwidth]{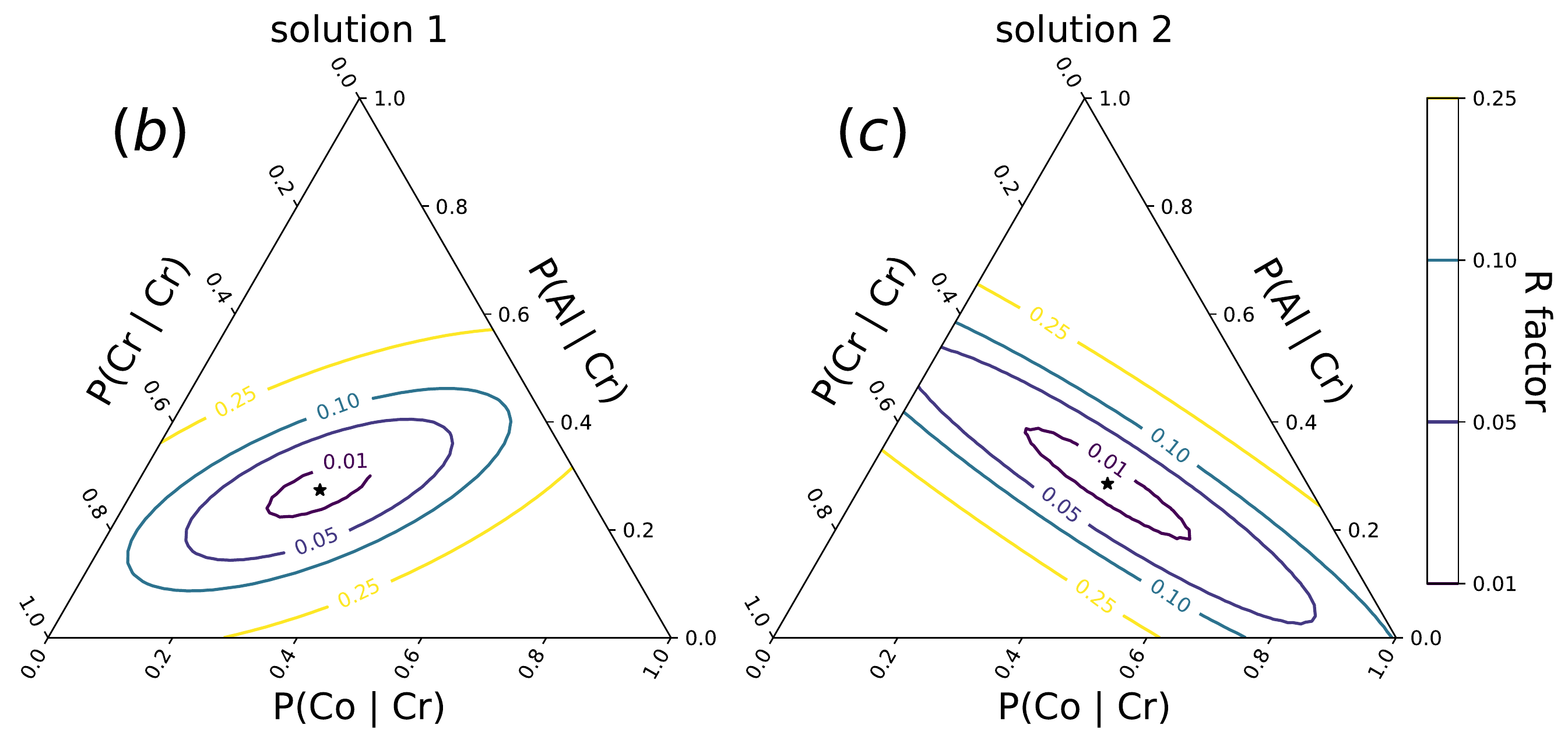}
    \caption{Interrogation of the error surface around solution 1 and 2 from Fig. \ref{fig:nonidentifiable}.
    (a): The lower panel shows the R factor along a trajectory consisting of a linear interpolation of all fit parameters between solution 1 and solution 2 (0 and 1 on the relative model coordinate respectively).  The upper insets show the fits at various points, indicated by  circles.  In these plots the black is the data and the colored lines are the fits.  The x-axis is in \AA{}.  
    (b) \& (c): Illustrates the error surface, plotted as R factor, about solution 1 and 2, fixing their respective fit parameters other than the composition of the \nth{1} NN coordination shell.}
    \label{fig:RdelE}

\end{figure}

We further interrogate this example by interpolating in parameter space between the two solutions.
Fig. \ref{fig:RdelE} (a) plots the R factor along this model interpolation trajectory with 5 selected model illustrations.
This shows that solutions 1 and 2 are distinct local minima and the fitting quality is degraded along the model interpolation coordinate.
The source of degradation is the exchange of first shell Cr and Co: The Cr-Cr bonds shift to longer distances with lower MSRD and lower occupancy, and the Cr-Co bonds trend in the opposite direction.
The EXAFS signals for these scattering paths phase shift in opposite directions along the interpolation trajectory, resulting in the positive amplitude deviation in the first shell when they contribute exactly in phase near the maximum of the R factor curve.
The change in other model parameters is negligible.

Fig. \ref{fig:RdelE} (b) and (c) show the sensitivity of the model to the first shell coordination variables around the two solutions, holding all other parameters fixed.
While Fig. \ref{fig:nonidentifiable} shows that these are distinct minima, it is notable that the two solutions show quite different correlations between coordination variables.
Low Z contrast between scattering species could explain the high correlation between the Cr-Cr and Cr-Co coordination variables in solution 1, while solution 2 counterintuitively shows high correlation between Cr-Al and Cr-Co scatterers that should be easily distinguishable.

\subsection{Unfavorable elemental contrast}\label{subsec:contrast}
While technically the design space of CCAs can include almost any element, frequently these elements are adjacent on the periodic table. 
For example, the quintessential light-weight CCA FeCoNiCrAl contains four first row transition metals plus Al\cite{sreeramagiri2021effect,gorsse2018high,sur2023high}.  
Similarly, many refractory alloys come from a 3x3 block of elements in the periodic table, so any combination of more than three refractories must contain at least one set of neighboring elements\cite{Lo2022,gorsse2018high}. 
These neighboring elements make EXAFS analysis much more difficult for two reasons.  

First, directly neighboring elements in some cases have absorption edges that are very close.
It is generally not possible to continue collecting EXAFS signal from one element above the absorption edge of another  element that is contained in the sample contains.
These cases limit the $k$ range of the data that can be collected and thus the information content of the data.   

The second issues with adjacent elements is the minimal difference in their photoelectron scattering cross-section.  
For this reason $f(k)$, the scattering amplitude, will show minimal contrast between neighboring elements. 
Because of their similar electronic configurations $\delta_\gamma$, the phase shift, will also show minimal variation between neighboring elements.
This is the principal reason for the extreme similarity between the scattering signal of the Fe, Co, and Ni species in Figure~\ref{fig:pathexpans}.

While there are various ``rule-of-thumb'' heuristics, most suggest that elements within 5 to 7 atomic numbers of each other cannot be distinguished in EXAFS\cite{Calvin2013}.  
However, distinguishing between scattering species of similar atomic number is not impossible.
Though they are highly similar, there is some contrast in   $f(k)$ and $\delta_\gamma$ that can be leveraged for fitting, particularly in near but not adjacent species.
More useful for speciation is the fact that different species have different bonding which changes both the associated MSRD and the bond length.  
Unfortunately, the high degree of parameter degeneracy discussed in Section~\ref{sec:degeneracy} makes isolating these effects difficult. 

One approach commonly used to mitigate this effect is to create a hybrid species of two or more elements\cite{zhang2017,sur2023high}, which are in turn represented by either one of the elements directly or an average species. 
The advantage of this approach is that it greatly reduces the number of free parameters in the model. 
However, there are  drawbacks.  
First, although neighboring elements may have similar electron scattering cross sections, the atomic radii of the elements may vary dramatically making the hybrid species non-representative.
Zr and Nb, for example, are neighboring elements commonly used in refractory alloys that have greatly differing metallic radii. 
A second drawback is that this approach reduces the scientific knowledge that can be gained from the fit and increases the difficulty of multi-edge fitting.
For instance in a FeCrAl alloy, if a fit is performed with a hybrid Fe-Cr species, then the specific occupancy of Cr in a shell about any of the scatters cannot be directly extracted from the model.
That being said, since there are only $N-1$ independent WC parameters for a single phase material,  application of the formulae in \textcite{de1971number} with a sufficient number of WC parameters from various edges and element combinations can yield these missing WC parameters.
If there is an insufficient number of values to fully determine the system of SRO parameters, these constraints can still provide bounds on the undetermined SRO parameters.

\subsection{Structural complexity and long-range order}\label{sec:LRO}
While CCAs are often idealized as chemically disordered solid solutions, the broader category of CCAs that we discuss here include alloys that more resemble traditional intermetallics, including alloys such as Heusler, half-Heusler, and Laves compounds.
In these alloys with long range order, there are distinct Wyckoff sites in the unit cell, multiplying the number of scattering paths required in the EXAFS path expansion model.
Consider a hypothetical non-stoichiometric ternary alloy \ce{A_xB_yC_z} with a B2 (CsCl) crystal structure, which has two distinct sites: an $a$ site at $(0,0,0)$ and a $b$ site at $(\frac{1}{2},\frac{1}{2},\frac{1}{2})$. 
In this alloy the \ce{A} atoms predominantly reside on the $a$ site, the \ce{B} atoms predominantly reside on the $b$ site, and the \ce{C} atoms are distributed between the two sites.
There are two crystallographically distinct local environments for \ce{C} atoms: those on the $a$ site, will mostly have \ce{B} atoms in their first neighbor shell, along with their associated bonding characteristics, while \ce{C} atoms on the $b$ site will mostly be surrounded by \ce{A} atoms.
% If one used the solid solution structural model with a single absorbing site for analysis of \ce{C} edge EXAFS for this alloy, the coordination numbers and potentially bond lengths and MSRD parameters would be sub
Standard EXAFS analysis, as described above, has an underlying assumption that bond length distributions are Gaussian.
When averaging the environments of the \ce{C} atoms across the two sites this assumption may be violated, particularly if the different sites have different neighbor distances.
Proper analysis requires separate \ce{C} absorption contributions for the two sites, doubling the number of fitting parameters associated with that element.

It seems unlikely that EXAFS analysis of alloys with partial LRO will be tractable without a good method for simultaneously quantifying the degree of long range order, especially if only a subset of species participate in the superlattice structure.
For example, the authors of reference~\cite{Cui2022} use HAADF-STEM to examine which elements participate in the sublattice ordering in an octonary alloy with both short and long range order.

\subsection{Compositional heterogeneity, second phase formation, and oxidation}\label{sec:secondphase}
% how to capitalize on challenges of EXAFS
% highly non-stoichiometric
% \begin{figure}[h!tbp]
%     \centering
%     \includegraphics[width=\textwidth]{complexity-discussion-starter.pdf}
%     \caption{DRAFT: ordered precipitates, compositional heterogeneity and grain boundary segregation, surface oxide structure and composition.}
%     \label{fig:heterogeneity}
% \end{figure}

While many CCA studies focus on or make simplifying assumptions limiting analysis to single phase and homogeneous material, in practice local heterogeneity and structural diversity is an important feature of these alloy systems.  
These inhomogeneities can take many forms, including secondary phase formation \cite{Jensen2016} %(Figure~\ref{fig:heterogeneity}a~\cite{Jensen2016})
, continuous composition gradients near grain boundaries and other microstructure features \cite{Mantha2021}
%(Figure~\ref{fig:heterogeneity}b~\cite{Mantha2021})
, and surface oxide formation \cite{Lo2022}
%(Figure~\ref{fig:heterogeneity}c~\cite{Lo2022})
.
Similarly SRO may vary spatially, with or without the presence of nominal composition variation. 
Each of these types of spatial inhomogeneities pose challenges for quantitative EXAFS analysis of SRO.

When there are discrete secondary phases, EXAFS measures the ensemble average of the environments of the absorbing atoms in the parent and any secondary phases. 
While the EXAFS decomposes into a simple sum of the contribution from each phase, in practice separating the contributions through explicit multiphase modeling is challenging and is not often attempted; the additional structural phase dramatically increases the number of parameters without increasing the available information.
For attempting such a fit, reliable estimation of the phase fractions and compositions of the constituent phases is critical.
However, supplementary measurement of these quantities presents yet another challenging measurement, analysis, and modeling problem.
This may present substantial difficulty if there are nanoscale precipitates that may require transmission electron microscopy techniques over more scalable probes like X-ray diffraction and scanning electron microscopy techniques.

In the case of localized compositional inhomogeneity, standard EXAFS at best can provide an ensemble average of the entire probed volume, which may be much larger than the length scale of the composition or structural fluctuations.  
However, as mentioned above standard EXAFS analysis assumes that bond lengths will have a Gaussian distribution. 
For such a continually varying system, it is not possible to know \textit{a priori} if these distributions will remain Gaussian.
Even in the case they are Gaussian, they are likely to be much broader than a homogeneous material.  
Without the ability to rely on modeling the MSRD as a narrow Gaussian, the analysis problem becomes much more complex.

In some CCAs, particularly refractory alloys, avoiding oxygen inclusion in the sample is nearly impossible.
%In fact, exclusion of oxygen may not even be desirable; preliminary evidence suggests that oxygen may drive SRO phenomena in CCAs, providing an important potential alloy design tool that must be understood.
Interstitial oxygen (or nitrogen) may be present and secondary oxide or nitride phases can form, typically leading to compositional inhomogeneity.   
Typical metal-oxygen bonds are shorter than metal-metal bonds, so the first metal-oxygen scattering paths contribute to the R-space EXAFS as a peak at somewhat shorter distance than the typical first metal-metal shell. 
While these peaks are often separable from the first metal shell through Fourier windowing, caution is warranted when neglecting the multiple scattering paths involving the first metal neighbor and an oxygen atom.
Furthermore, the nearest metal-oxygen scattering contribution is concentrated in a portion of the EXAFS that is critical for robust background estimation~\cite{newville1995analysis}.
This increases the difficulty of selecting good (and transferable) background estimation settings and can potentially bias estimates of coordination number variables.

At low levels, oxygen will typically cluster to the interstitial sites surrounding the most gettering species, and the oxygen-metal multiple scattering paths can likely be safely ignored. 
However, at higher levels this can begin to interfere with the fitting.
If discrete oxides form, the second phase must be explicitly modeled to account for the contribution of metal absorbing sites in the oxide to the EXAFS signal, as discussed above in the context of intermetallic compounds.

\section{Possible ways forward}\label{sec:approach}
We will need innovation on multiple fronts in order to mitigate the challenging aspects of EXAFS analysis of scientifically and technologically interesting CCA systems.
We must think carefully about how to improve the quality of the data we collect, increase the amount of information we bring to the quantitative analysis table, and develop more powerful spatially-resolved and multimodal analytical microscopy techniques.

\subsection{Improved data quality}
Proper data collection and its resultant quality is important for producing usable EXAFS analysis.
Best practices in this area are generally well known~\cite{Calvin2002}, but there are opportunities for  improvement.
For tasks such as extracting WC parameters, where coordination number is critical, getting proper peak intensities is necessary.
These peaks can be weakened by self absorption in thick, concentrated samples measured in fluorescence mode.
While this is readily avoidable in CCAs synthesized as thin films, bulk samples must be thinned to a level that can be a challenge for conventional metallurgical sample preparation over areas larger than beam size for standard bulk EXAFS analysis.  Collection of standards can also help in this regard \cite{newville1995analysis}.

While the Nyquist criterion tells us that spectra with larger measured range of $k$ space contain more information, the $k$ range of the scan is not the only parameter that affects data quality. 
EXAFS data are commonly collected with non-uniform energy discretization, typically using a much finer energy resolution and integration time in the absorption edge than in the extended EXAFS region, with intermediate energy resolution in the pre-edge part of the scan.
Both the density and collection time can be adjusted, not just for these common scan segments, but potentially dynamically adjusted based on the characteristics of the data as it is being collected \cite{leshchev2022inner}.
Given a fixed measurement budget, optimal allocation of measurement time across the full absorption spectrum for robust and efficient quantitative analysis is an open question.

\subsubsection{Spatial mapping}\label{subsec:exelfs}
In multiphase or inhomogeneous systems, robust quantification of SRO would likely require localized EXAFS data that captures a single structure and composition regions of the sample, often on the nano-scale.
There are two approaches to collecting these data along with accompanying compositional data (and potentially structural data from diffraction), both still largely in development.

The first approach is to use hard x-ray nanoprobe~\cite{mino2018materials}.  
Beams on the order of 100 nm can be readily generated using reflective, and thus achromatic, optics such as KB (Kirkpatrick–Baez) mirror pairs and have been demonstrated for measuring EXAFS \cite{martinez2014exploring}. 
However, the zone plates and other Fresnel optics and multilayer Laue optics that rely on refraction or diffraction typically used to generate beams smaller than this are inherently chromatic optics, which changes the focal length as a function of energy.  The chromatic effects on beam size  along with monochromator induced beam movement and flux limitations make extended EXAFS measurements at this sample size largely intractable with current technology \cite{pattammattel2020high,mino2018materials}.  
To the extent that these measurements are tractable, one advantage is that they are amenable to fluorescence mode XAS with relatively thick samples.  

The other option is to use a scanning transmission electron microscope (STEM) equivalent to EXAFS, known as extended energy loss fine structure (EXELFS) to provide a dramatic improvement in spatial resolution\cite{hart2023revealing}.
EXELFS, like EXAFS, provides the same information regarding the local atomic environment, but with nanometer spatial resolution and an energy resolution down to $< 1$ eV. 
Until recently, application of EXELFS to study SRO has been limited due to low energy resolution and limited energy range to access relevant absorption edges.
\textcite{hart2019synchrotron} resolve these issues through state of the art direct detection EELS to enable spatial SRO mapping with nanometer resolution.
%Recently,~\textcite{hart2019synchrotron} enabled rapid high resolution 
% Previously, the application of EXELFS to study SRO was very limited due to the poor quality of EELS data, however, \textcite{hart2019synchrotron}, using state-of-the-art EELS instrumentation
% %\cite{hart2017direct}
% , recently developed  EXELFS, and demonstrating its utility by measuring local order in a Cu-Zr-Ni-Al bulk metallic glass (BMG). 
They demonstrate by locally measuring the SRO in a Cu-Zr-Ni-Al bulk metallic glass (BMG), as a function of distance from the interface of the BMG film to a Ni layer. 
They found a positive Warren-Cowley parameter $\alpha$\textsubscript{Ni-Zr} (Eq.~\ref{eq:WCalpha}) across the entire BMG width, indicating a preference for Ni and Cu clustering.
An additional advantage of EXELFS is that the sample can be further analyzed by other high resolution TEM technique to fully understand the related properties. 

One drawback of these high spatial resolution measurements is that they require large measurement time and only provide information for a localized area. 
For a single sample the measured area may not be representative and the substantial sample preparation investment makes systematic high-throughput EXELFS studies intractable. 
One possible path forward is to use preliminary detailed EXELFS spatial mapping to characterize local structural and chemical distributions.
This characterization can inform the model specification and optimization for quantitative or semi-quantitative analysis of high throughput ``bulk'' EXAFS measurements.
Finally, the high throughput bulk EXAFS analysis may identify selected areas for subsequent spatially-resolved EXELFS analysis.
In principle, the spatially-resolved EXELFS data and the bulk EXAFS measurements can be modeled jointly as well.
% bravel: I think this is an AMAZING idea.  Imagine finding regions of pure phase material in the EXELFS and using that to inform the EXAFS analysis of the mixed phase material.  Even having a subset of phases in EXELFS could go a long way towards making a complex EXAFS problem tractable.

\subsection{Increasing information with larger datasets and multimodal characterization}
Because of the complications described above, the likely path to robust extraction of useful SRO knowledge from EXAFS almost certainly involves the injection of additional information to reduce, or at least constrain, the dimensionality of the fit.  
Potential information sources include (justifiable) simplifying assumptions, joint modeling of related EXAFS data, information fusion through multimodal characterization and analysis, and incorporation of data and knowledge from supplementary information streams such as computational thermodynamics and atomistic simulation.

\subsubsection{Joint analysis of multiple edges}\label{subsec:multiedge}
As discussed in \S~\ref{subsec:complexity}, the degree of overparameterization can be reduced by coupling bond length and MSRD parameters across pairs of absorption edges through a symmmetry assumption.
However, imposing similar pairwise constraints on the coordination number variables in solid solutions is not so straightforward.
As discussed by \textcite{de1971number} the coordination numbers of pairs of species are related by the definition of conditional probability as well as simplex constraints, reducing $M^2 - M$ coordination parameters to $(M^2 - M) / 2$.
However, naive application of these coordination symmetry constraints in non-equiatomic solid solutions can result in invalid SRO matrices that imply negative coordination numbers.
Application of the quadratic constraints identified by de Fontaine can resolve this issue, 
but these constraints couple across pairs of components, complicating straightforward application of these constraints in nonlinear least squares model optimization, as outlined in our forthcoming paper~\cite{SROManifold}.

% The SRO assumption set is able to sufficiently reduce the model complexity for FCC solid solutions of up to order 4, but only if all absorption edges are modeled together.
% If even one edge is not available--a common occurrence due to the energy range limitations of beam lines--then more restrictive assumptions and/or additional sources of information are needed to make analysis tractable.

% The lower portion of the diagram represents the EXAFS pattern; for each $i$ in $I$ data-points, there exists a $y_i$ corresponding to the noiseless signal and an $\epsilon_i$ representing the noise, each of which in-turn effect the measured intensity, $y_i$, for that data-point.  For simplicity, not all variables or connections are shown. 

\subsubsection{Joint analysis of multiple samples}
There are also cases where data on related samples are collected simultaneously.
This may include samples with neighboring compositions, like combinatorial libraries \cite{maffettone2022selfdriving,JORESS2022353}, or the same sample over multiple temperatures.
In these cases, at least within a single phase, we can assume the fitting parameters will vary smoothly.
For instance, we would expect $\sigma^2$ parameters to increase monotonically with temperature, which is routinely modeled with the correlated Debye model for joint analysis of EXAFS collected over a temperature series~\cite{Dalba1997}.
To constrain the fit and exchange information between them, we can perform a global fit on all the samples using a model of the type illustrated in Fig.~\ref{fig:model-diagrams}.  
In this hierarchical model the EXAFS parameters for each sample are drawn deterministically from a smooth function, \textit{e.g.,} a spline or Gaussian Process~\cite{williams2006gaussian}.  
The fit in this case is performed by optimizing the parameters of those functions rather than the individual parameters.  

\subsubsection{Data fusion and multimodal analysis}
Above, in \S\ref{subsec:exelfs}, we describe an approach to deal with inhomogeneous samples by combining EXELFS and EXAFS data. 
Beyond the problem of sample inhomogeneity, data augmentation can be useful to inform and bias EXAFS modeling. 
It is possible to augment EXAFS data with both computational and experimentally obtained values.
For instance, data obtained from pair distribution measurements (PDF) can be added to constrain the fit\cite{maffettone2022selfdriving}. 
In the example shown in Fig. \ref{fig:nonidentifiable}, PDF may be able to provide a set of bond lengths, identifying one of the two models as more likely.
We note however that x-ray scattering cross sections and electron scattering cross sections are similar so only a small amount of additional information is available from x-ray PDF when there is low Z contrast.
Neutron cross-sections, conversely, are very different as a function of Z so that neutron based PDF can provide information complementary to EXAFS.
Unfortunately, the sample preparation required for these two techniques is quite different, and is not amenable to current high throughput material synthesis capabilities.  

Computational methods, especially atomistic simulation such as density functional theory (DFT) calculations and  molecular dynamics (MD) simulations, can also provide valuable data to aid in fitting the EXAFS data, particularly in helping define starting parameters or identifying plausible physical constraints.
For example, MD simulations can be used to parameterize bond length parameters in highly disordered systems, particularly those that require the bond length histogram method~\cite{Price2012}.
For high throughput experimental studies, it may be helpful to base  model specification and initial parameter guesses on average bond lengths, MSRDs, and coordination numbers obtained from hybrid Monte Carlo / Molecular Dynamics calculations, as in~\cite{farkas2018model}.
A variation of this approach has been the focus of recent efforts in ML-driven characterization~\cite{Timoshenko2020,Schmeide2021}.
This approach uses large sets of Molecular Dynamics calculations (often enabled for complex alloy systems by machine learning interatomic potentials) to directly predict the parameters of the structural model from simulated EXAFS data.
One drawback of this general approach is that it introduces yet another machine learning research problem: rapid development of robust multi-component interatomic potentials~\cite{deringer2019machine}.

Similarly DFT, potentially coupled with statistical thermodynamics methods, can directly provide strong prior knowledge on the relative stability of compounds and alloys~\cite{Hegde2020}.
This could both accelerate and improve the accuracy of model search and specification for complicated multiphase alloys, particularly when multiple potential intermetallic phases could form.
Furthermore, DFT estimates of the interaction energies between different atomic species can inform theoretical predictions of the relative ordering preferences between elemental species~\cite{Rao2022,Rowlands2006,FernndezCaballero2017}, which can either constrain or help initialize coordination number estimates in EXAFS analysis.

%\textbf{Work with Wolverton group to follow \cite{Wang2018} to get clarity on oxygen interstitials?}

\subsection{What simplifying assumptions are defensible?}
Commonly, overparameterization of EXAFS models is further mitigated by introduction of physically defensible constraints between parameters;
Chapter 14 of \textit{XAFS for Everyone}~\cite{Calvin2013} discusses in detail a variety of physical (\textit{e.g.,} known long range crystal structure) or simplifying (\textit{e.g.,} arbitrary constraints on bond lengths) constraints~\cite{Calvin2013}.
Given the complexity of CCAs and the importance of local lattice distortion, it is unclear to what extent many of the commonly used assumptions are safe to apply.

% Note, our pseudobinary model is similar to that of~\cite{Zhang2017}, but they fix coordination variables to equiatomic distribution.

% \cite{Lilensten2022} uses symmetric bond lengths and coordination numbers, but a single MSRD parameter.

% \cite{Tan2023} this paper constrains all the MSRD parameters to be equal, and fits coordination numbers and bond lengths. They also set Nb and Zr edge energy shifts equal and fix $S_0^2$ to a value they got from fitting Zr and Nb standards.

For example, a simplifying assumption that is frequently employed is that all atomic pairs have the same MSRD.  However, in alloys with large degrees of static lattice distortion, care must be taken in how MSRDs and bond lengths are parameterized in a fit.
The effects of static and thermal disorder are quite difficult to disentangle in EXAFS analysis.
Doing so requires high data quality over an extended energy range and benefits from measurement at multiple temperatures. 
%% (BR: I am giving a stab at rewriting the rest of this sentence) 
%% assuming common bond lengths and/or MSRD parameters between all pairs of species may severely bias the fitted coordination numbers because all three variables can shift the phase of the modeled EXAFS signal of the alloy.
Alternatively, it may be appropriate in some cases to fit $M$ metallic radii instead of an $M+(M^2 - M) / 2$ symmetric matrix of first shell bond lengths. This gets more complicated for high level NN shells, including the \nth{2} NN shell in BCC.  We tabulate the total number of free parameters when applying this constraint in addition to the symmetry and SRO constraints to fits of FCC materials in the \nth{9} and \nth{10} columns of Table \ref{tab:overparameterization}, respectively.
Another approach is to constrain the model to a fully random SRO distribution and qualitatively infer the ordering tendencies based on the fitted bond length distributions~\cite{zhang2017}.
It is also common to fit $S_0^2$ to elemental standards and assume the values transfer to the crystal structures and sample morphologies being measured, which may differ substantially from those of the elemental standards.  This simple approach to the amplitude reduction factor can be improved by consideration of many-body effects. \cite{B926434E}
Considering the extreme overparameterization of BCC and multiphase CCA structural models, analysis may not be tractable without some kinds of simplifying assumptions. The CCA community would benefit from a systematic benchmark study to firmly establish criteria for when different assumptions may be acceptable given the scientific goals.

\subsection{Multiple scattering and higher shell paths}\label{subsec:highR}
To this point we have predominately focused on fitting of first NN shell (or in the case of BCC first and second NN shells) as including additional shells greatly increases the complexity of the model.
Further, EXAFS in these alloys tends to have minimal spectral weight at higher R values.
To the extent it is possible to include these multiple scattering paths, however, there may be benefits.
First, higher shell fits have greater ability to decorrelate $S^2_0$ and $N$. 
Using higher R ranges also increases the number of parameters according to the Nyquist criteria. 
Finally, the multiple scattering and higher shell parameters can further constrain the fit by adding beneficial bias through structural assumptions.

\subsection{Other computational approaches}  
% \textbf{note: add a topic on increasing computation? need discuss reverse Monte Carlo, and ML methods can maybe go under this heading...}
In addition to data fusion methods combining physics-based computational modeling with x-ray absorption spectroscopy (XAS) described above, ML has the potential to be a powerful tool for EXAFS analysis.
There are various examples of   such frameworks in microscopy and spectroscopy, providing inference from computationally generated structures \cite{Timoshenko2020,Unruh2022}. 
Many studies approach the EXAFS fitting problem from a black box optimization perspective using methods such as genetic algorithms to perform quantitative analysis \cite{Terry2021,Martini2021,kido2020problems}.
While these methods can more efficiently search high-dimensional parameter spaces, they do not alleviate the extreme degeneracy of the structural models needed for robust and complete analysis of solid solutions and multiphase CCAs.
% While providing improved tools for fitting EXAFS data, these methods don’t add further information than what is presented within the provided data.
In addition to directly improving quantitative analyses, ML offers powerful non-linear inference capabilities for accelerated acquisition and model selection for spectral data.
Autoencoders have been demonstrated in EELS and XAS analysis in extracting a structurally relevant latent feature space \cite{Pate2021, Kalinin2021, Tetef2021}, enabling rapid classification and reduced-order modeling of properties influenced by material structure.
For EXAFS, this approach could be used to rapidly identify which structural models and corresponding assumptions may be appropriate for a given observation.
Beyond unsupervised autoencoder approaches, physics informed ML frameworks add a further level of information extraction since models are trained using computationally generated data.
Optimizing further to incorporate multiple modalities will build out highly correlated analysis capabilities beyond what is present in the EXAFS signal alone, and improve ML inference beyond the scope of a single characterization method \cite{maffettone2023self}.

\section{Summary}
Robust and efficient quantitative EXAFS analysis can enable the SRO measurements that are required to elucidate the mechanistic underpinnings of the exciting properties that CCAs offer.
There are many challenges described here in achieving this.  
Fortunately there are still many underexplored avenues that may lead to improved measurement power.  While we hope that this work serves both as a guide and a call-to-action for the CCA research community to improve the quality and information content of EXAFS fits, we also point out that currently most of the approaches described here are not simple or tractable to implement using common analysis tools such as Artemis\cite{ravel2005athena}, and broad scale adoption will also require efforts by the EXAFS community to adapt software to this complex but impactful challenge.

% Taking some notes for the figures
% \section*{Scratch}
% Additional sources of information?

% \begin{itemize}
%     \item reduce SRO parameter range based on high throughput DFT relative energies with Wolverton group
%     \item bond lengths and/or variances from atomistic simulation?
%     \item joint analysis of X-ray PDF or XRD
%     \item hierarchical modeling of multiple related spectra to pool information
%     \item limited neutron scattering, then information propagation to high throughput EXAFS?
% \end{itemize}
% \subsection*{Figure drafting}
% Figures?
% \begin{itemize}
%     \item Complexity scale: path expansion
%     \item Multi-scale figure, 2-phase regions, mat sci concepts 
%     \item parameter spread, degeneracies, fitting maps?,
%     \item plate diagram?
%     \item spatially resolved EXELFS
% \end{itemize}

% \section*{Contribution Statement}
% HJ, BD, and MLT conceptualized the manuscript and framed the goals.
% HJ and BD performed the formal analysis in Sec~\ref{sec:challenges} and created the visualizations.
% BD performed the case studies in Figures~\ref{fig:nonidentifiable} and ~\ref{fig:RdelE}
% HJ, BD, JH, EA, and DS wrote the initial draft of the manuscript.
% HJ, BD, BR, JHS, and MLT performed critical review and revision.
% MLT acquired financial support and material resources.
% BR provided experimental resources for EXAFS data collection at the 6-BM beamline.

\section*{Acknowledgements}
This research used resources at the 6-BM beamline of the National Synchrotron Light Source II, a U.S. Department of Energy (DOE) Office of Science User Facility operated for the DOE Office of Science by Brookhaven National Laboratory under Contract No. DE-SC0012704.
The authors gratefully acknowledge partial funding from the Office of Naval Research (ONR) through the Multidisciplinary University Research Initiative (MURI) program (award \#: N00014-20-1-2368) with program manager Dave Shifler.
%The authors acknowledge Prof. Jason Hattrick-Simpers for his careful reading of our manuscript and his insightful questions/suggestions during its revisions.

\section*{Declarations}
%\subsection*{Declaration of Interests}
The authors declare no competing interests.

The opinions, recommendations, findings, and conclusions of this work do not necessarily reflect the views or policies of NIST or the United States Government.

%\printbibliography
\bibliography{references}
\end{document}